\def \be {\begin{equation}}
\def \ee {\end{equation}}
\def \bea {\begin{eqnarray}}
\def \eea {\end{eqnarray}}

\def \dels {\partial\kern-.5em / \kern.5em}
\def \As {{A\kern-.5em / \kern.5em}}
\def \Ds {D\kern-.7em / \kern.5em}

\def \Om {\Omega}

\newcommand{\hoch}[1]{$\, ^{#1}$}

\documentclass[12pt]{article}

\begin{document}
\begin{titlepage}

\begin{center}
\hfill hep-th/0101058\\
\vskip .5in

\textbf{\large Stringy Spacetime Uncertainty
as an Alternative to Inflation
}

\vskip .5in
{\large Je-An Gu\hoch{1}, Pei-Ming Ho\hoch{1}
and Sanjaye Ramgoolam\hoch{2}}
\vskip 15pt

\begin{itemize}
\item[$^1$] {\small \em Department of Physics,
National Taiwan University, Taipei 106, Taiwan, R.O.C.}
\item[$^2$] {\small \em Department of Physics,
Brown University, Providence, RI 02912, USA}
\end{itemize}

\vskip .2in
\sffamily{
wyhwange@phys.ntu.edu.tw\\
pmho@phys.ntu.edu.tw\\
ramgosk@het.brown.edu}

\vspace{60pt}
\end{center}
\begin{abstract}

In this paper we point out that
the spacetime uncertainty relation proposed
for string theory has strong cosmological implications
that can solve the flatness problem and the horizon problem
without the need of inflation.
We make minimal assumptions about the very early universe.

\end{abstract}
\end{titlepage}
\setcounter{footnote}{0}

Inflation is becoming one of the most
important ingredients in modern cosmology.
It provides solutions to the flatness problem,
the horizon problem and the monopole problem.
It also provides a framework to construct models
that can explain existing experimental
observations, in particular
the spectrum of cosmic microwave background radiation
(CMB) anisotropies.
However, inflationary models are frequently
involved with physics at very high energy scales,
such as the GUT scale or even the Planck scale,
far above the energy scale which has been
experimentally confirmed ( see for example \cite{BM} ).   
In the light of the brane world scenario \cite{AH},
it is possible that the fundamental scale
of quantum gravity, or the scale of string theory,
is much smaller than the 3+1 dimensional Planck scale.
This means that certain new physics
very different from conventional field theory
may play a crucial role in the early universe,
and thus our current understanding of cosmology
may have to be greatly modified accordingly.
The fact that the physics at very short distances
can be amplified to observable cosmological effects
is in fact one of the reasons why cosmology 
can be a crucial arena for testing  string theory. 

In this paper we point out such a possibility.
We find that the spacetime uncertainty relation
appearing in string theory have
cosmological implications so significant that
it may even resolve two of the above-mentioned
puzzles without inflation.
Remarkably, as we will see below,
even if the energy scale associated with
this spacetime uncertainty relation
is as high as the Planck scale,
it may still have equally strong cosmological implications.

In string theory it is proposed \cite{Yoneya1,Yoneya2}
that there is a spactime uncertainty relation
\be \label{ur}
\Delta x\Delta t\geq l_s^2,
\ee
where $\Delta x$ and $\Delta t$ are the uncertainties
in the measurement of length in any spatial direction
and that of time, and $l_s$ is the string length scale.
This relation is of a stringy origin.
The simplest argument for it is the following.
In quantum mechanics we have the uncertainty relation
\be
\Delta E \Delta t > 1.
\ee
For a string the uncertainty $\Delta E$ in energy
is related to an uncertainty $\Delta x$ in its spatial extension
\be
\Delta E\sim T_s \Delta x,
\ee
where $T_s=l_s^{-2}$ is the tension of the string.
Combining the two relations above we obtain (\ref{ur}).
It follows that when we use strings to probe
the spacetime geometry, the precision in our measurement
is limited by this uncertainty relation.

Note that in string theory, the property of spacetime
is defined by the dynamics of the strings.
The meaning of $\Delta x$ in (\ref{ur}) is different
from that of $\Delta x$ in Heisenberg's uncertainty relation.
In quantum mechanics, each measurement of $x$
gives a definite value of $x$.
Heisenberg's relation means that the distribution of
possible values of $x$ is of scale $\Delta x$.
On the other hand, for the spacetime uncertainty relation,
what $\Delta x$ means is that at the string scale,
one can no longer think of the spacetime as a classical manifold.
Rather, the spacetime has to be described by
certain noncommutative geometry or quantum geometry,
for which it doesn't make sense to talk about
two points separated by a distance smaller than $\Delta x$.
For a field theory living on such a quantum space,
it is therefore impossible to associate independent
physical degrees of freedom within
a range smaller than $\Delta x$.
The meaning of $\Delta x$ in (\ref{ur}) is thus
more fundamental than that in Heisenberg's relation.

If the string length scale $l_s$ is smaller than
the scale currently accessible in experiments,
say, smaller than $(TeV)^{-1}$,
we might expect that we need to build more powerful
colliders in order to see the effects of
the uncertainty relation.
However, as we will see below,
it can leave clear cosmological imprint to be
readily observed in the sky.

To fix our notation, recall that in standard cosmology,
the Friedmann-Robertson-Walker (FRW) metric of the universe is
\be \label{metric}
ds^2=dt^2-R^2(t)\left(\frac{dr^2}{1-kr^2}+r^2 d\Om\right),
\ee
where $k=0,1,$ or $-1$,
corresponding to a flat, closed or open universe.
The metric takes this form whenever
the universe is isotropic and homogeneous.
For $k=\pm 1$, $R(t)$ is also the radius of curvature
(up to a constant numerical factor).
For $0<t<t_{EQ}$, the universe is radiation dominated,
and $R(t)\propto t^{1/2}$.
For $t_{EQ}<t$, the universe is matter dominated,
and $R(t)\propto t^{2/3}$.
At $t=t_{EQ}\sim 10^{10} sec$,
the radiation density equals the matter density.
The Hubble radius $l_H$ and the particle horizon $d_H$
are roughly the same $l_H(t)\sim d_H(t)\sim t$,
up to an overall constant factor of order one.
As a classical solution to the Einstein equation
for four dimensional (4D) spacetime,
we expect that the FRW metric be modified
when the temperature exceeds the energy scale
of new physics, say the string scale.
However, the qualitative discussion below
are robust to changes in the exact solution,
as long as the universe starts with a big bang
at a certain instant of time.

Eq.(\ref{metric}) describes the evolution of the universe
for sufficiently late times.
Near $t=0$, the universe may have
a completely different description,
involving stringy effects that patch to an
earlier contracting phase \cite{Ven},
or a gauge theoretical dual,
or a phase dominated by D-branes \cite{mag},
or a spacetime ``foam'' of virtual black holes \cite{per}.
We assume it makes sense to define the extent in time,
from $t=0$ to $t=l_*$, of such an early phase up to
an uncertainty $\Delta t$.
Clearly $\Delta t<l_*$.
This relation together with (\ref{ur})
implies that
\be \label{dx}
\Delta x>\frac{l_s^2}{l_*}.
\ee
As a fundamental uncertainty in the measurement
of length, $\Delta x$ is the lower bound
on essentially any physical length variable.

In addition to the spacetime uncertainty relation,
it was often thought that another uncertainty is true
which claims that $\Delta x$ and $\Delta t$ should
by themselves always be greater than $l_s$.
This claim is now believed to be incorrect, after 
developments showing that $D0$-branes can probe very short 
distances \cite{DKPS}.
We are not assuming such fundamental uncertainties
separately on time and space, in agreement with \cite{Yoneya2}.
In our application, $l_*$ will depend on
the nature of the evolution of the early universe.
Possible candidates include $l_p$,
the 4D Planck scale, or $l_{11}$,
the 11D Planck scale.
It is important that it is different from $l_s$,
and further, for our application to cosmology,
it is taken to be less than $l_s$.
For simplicity we first focus on the case
$l_*=0$.

The flatness problem is to explain why 
the universe is observed to be so close to flat,
or equivalently, why the energy density is
so close to the critical density.
Another way to describe the problem is to
note that the radius of curvature $R(t)$ is about $10^{30}$
times larger than the characteristic length scale
$l_H$ or $d_H$ at Planck time $t_P\sim 10^{-44} sec$.
The uncertainty relation solves the flatness problem
in a very simple way.
At any time $t$, the radius of curvature $R(t)$ has
to be larger than $\Delta x$.
Since $t>\Delta t$, it follows from the uncertainty relation that
$R(t)>l_s^2/t$ at any time.
As we take $t$ all the way down to $t=0$,
we see that the radius of curvature has to be infinite,
and so the universe is completely flat at $t=0$.
The 
metric (\ref{metric}) then implies that $k=0$,
and the universe is flat at all times.
Note that
$R(t)$ is now only a scaling factor
and is not a physical length scale
for the case $k=0$.
It does not have to satisfy $R(t)>\Delta x$.
This argument works equally well
no matter how small $l_s$ is,
as long as it is not zero.
In addition, it works for any dependence of $R(t)$ on $t$.

The spacetime uncertainty relation
also solves the horizon problem,
which is essentially asking why the CMB radiation
is isotropic
on the last scattering surface
while the particle horizon at the decoupling time
($t_d\sim 10^{12} sec$) can only explain
a very small fraction ($10^{-5}$) of the observed region.
According to (\ref{dx}),
contrary to the ordinary expectation that
at earlier times the fluctuations are more violent,
the universe is actually much smoother.
Although the physics in different horizons
are not causally related at $t_d$,
they still have strong correlation because
fluctuations of short wavelength can not exist
when time $t$ is small.
In fact, at $t=0$, the size of particle horizon is zero,
but the whole universe is uniform,
with total correlation between any two point
in space no matter how far they are separated.
In some sense, the uncertainty relation with $l_*=0$
dictates the universe to start with a very
peculiar initial state.
{}From the viewpoint of ordinary field theories,
such states would be considered almost impossible
to occur because they are so rare that they
constitute a set of measure zero \cite{CH}.

As an example of nonzero $l_*$, let
\be \label{ex}
l_s\sim (TeV)^{-1}, \quad
l_*\sim\l_p\sim(10^{19}GeV)^{-1}\sim 10^{-44}sec,
\ee
where $l_p$ denotes the 4D Planck length.
We choose $l_s$ to be of this value because
it is roughly the largest possible value
without contradiction to particle experiments.
Although $l_s$ enters the spacetime 
uncertainty principle since we expect stringy 
degrees of freedom to become relevant at that 
scale, $l_*$ has  been identified with $l_p$ 
since this is where one might expect a 
four dimensional description which ignores 
 the dynamics of the extra dimensions to break down 
completely. 
For this case (\ref{ex}),
one has $\Delta x\simeq 10^{-2} cm$ at $t=l_*$.

For the flatness problem,
using $R(l_*)>l_s^2/l_*$ we find
\be
\frac{R(l_p)}{d_H(l_p)}>\frac{l_s^2}{l_p^{1/2}l_*^{3/2}}.
\ee
It follows that this ratio is about $10^{32}$
for the example (\ref{ex}),
in agreement with the experimental bound ($10^{30}$).
For the horizon problem,
fluctuations can exist only at
a length scale larger than $l_s^2/l_*$ when $t=l_*$.
At decoupling time $t_d$,
this smoothness scale is amplified into the size of
\be
L\simeq\frac{l_s^2}{l_*}\frac{R(t_d)}{R(l_*)}.
\ee
Causal interactions will result in a correlation length
larger than $L$ by the size of the particle horizon at $t_d$,
but it is negligible compared to
the smoothness we need to account for.
If $l_*$ is sufficiently smaller than $l_s$,
it is possible to have $L$ large enough to
agree with CMB observations.
For the example (\ref{ex}),
$L$ is roughly the same as the size of today's horizon.
and suffices to account for the homogeneity
we observe in CMB today.
For fixed $l_s$, the smaller $l_*$ is,
the longer we will have to wait
until we see
comparable anisotropies
in CMB.
One can say that we are replacing the two problems
by the new hierarchy problem why $l_s/l_*$ is so large.
This may turn out to be a dynamical result of
the evolution of the early universe,
or an initial condition problem.

So far we have not addressed the problem of monopole
abundance. A treatment of this problem requires
an understanding of the thermal history of the universe,
and more details of the new physics at the string scale.
It cannot be solved by the uncertainty relation alone.
Nevertheless, roughly speaking, the uncertainty relation
forbids high momentum modes from appearing  at early times.
It makes it harder to generate a lot of heavy particles.
On the other hand, there are many moduli fields
in string theory which may show up as relics
for us to worry about, just like monopoles.

Another important issue is the energy density perturbation.
The primordial energy density perturbation is found to
have a scale invariant power spectrum,
and to satisfy the Gaussian distribution.
Just like for inflationary scenarios
one needs to construct specific models
to account for these experimental data,
here we need to first construct a model,
which obeys the uncertainty relation (\ref{ur}).

There are two obvious choices to realize (\ref{ur})
in a field theory.
One way is to consider a noncommutative spacetime
which satisfies some nontrivial commutative
relations from which (\ref{ur}) can be derived.
This may require some care
since  noncommutativity between time and space
variables is expected to  cause many difficulties in field theories,
such as the violation of causality and unitarity \cite{ts},
although it may be allowed in string theory \cite{BR}.
However non-commutativity in both space-space 
and space-time directions appears to be possible 
for decoupled theories \cite{AGM}. Non-commutativity 
in spatial as well as space-time directions
has also  been related to dual CFTs
in the context of the ADS/CFT 
correspondence \cite{JR, HRT, HL}.  
Generalizations of Riemannian geometry and general relativity
to noncommutative spaces have been proposed \cite{Connes,Ho},
but a better understanding is still needed.

Another possibility is to modify the canonical
commutation relation between $x$ and $p$,
such that from this relation one can derive
\be \label{dx1}
\Delta x>\frac{l_s^2}{t}.
\ee
The case of a time-invariant uncertainty for $\Delta x$
is considered in \cite{Kempf}.
It is straightforward to modify it to have
the uncertainty (\ref{dx1}).
The drawback of this approach is that we are not
realizing the fundamental uncertainty relation (\ref{ur})
but only a consequence of that.
Its merit is that it is very close
to a traditional formulation.
In this approach the Hilbert space consists of
fluctuations of all frequencies all the time,
but the Hamiltonian operator vanishes on
higher Fourier modes until the time when
$\Delta x$ is smaller than its wavelength.
In other words, those fluctuations violating
the uncertainty relation are spectators which
are decoupled from everything else.

An exploration of these possibilities
is left for future study \cite{prep}.
Here we shall only discuss what are the implications
of the observed properties of the energy density perturbations.
The fluctuation $\delta\rho$ in energy density
with comoving momentum $k$ has physical momentum
$k/R(t)$ at time $t$,
and thus is forbidden for $t<(kl_s^2/a)^{2/3}$
for $R(t)=at^{1/2}$.
It follows that fluctuations of different length scale
we see today emerge at different $t$.
The fluctuations of interest to us are
those of sizes $1 Mpc$ to $1000 Mpc$.
They emerge out of the uncertainty constraint
between $t\sim 10^{-42} sec$ and $10^{-40} sec$
for the example (\ref{ex}).
As in inflation, longer wavelengths appear earlier in time.

At first glance, the fact that fluctuations of
length scale smaller than $\Delta x(t)\sim l_s^2/t$
are forbidden may seem to violate causality.
One may argue that if someone
interferes with a field at a certain place,
the field will have to react simultaneously at
another place $\Delta x$ away from it.
This is however not a valid argument.
As we have tried to emphasize ealier,
the quantum geometry of spacetime is such that
it does not make sense to talk about
independent physical degrees of freedom at
two points within a range of $\Delta x$.
Any interference one can perform 
to a physical system is done over a region of
size $\Delta x$ simultaneously.

Our discussion of the horizon and flatness problems
has made minimal assumptions about the
very early universe, except for associating with it
a small uncertainty in time, in addition to
the stringy uncertainty principle.
We presented scenarios where these problems
are solved without any need for inflation,
but it is also possible to use this mechanism
in conjunction with inflation.

In the brane world scenario,
it is also of interest to consider
cosmological effects of noncommutative geometry
associated to the worldvolume of the brane,
as in ref.\cite{CGS}.
This possibility is also motivated from string theory,
where it was found that
a D-brane worldvolume becomes noncommutative
when the NS-NS B field background is turned on \cite{NC}.
It was also pointed out that even if
the vacuum expectation value of the B field vanishes,
its quantum fluctuations will result in
an effective uncertainty relation for
the measurement of spacetime coordinates
on the D-brane \cite{CHK}.
It would also be interesting to consider
the cosmological effects of these uncertainty relations.

\section*{Acknowledgment}

It is a pleasure to thank S. Alexander, R. Brandenberger,
H.-C, Cheng, B.-H. Lee, T. Wiseman and T. Yoneya for discussions. 
The work of P.M.H. is supported in part by
the CosPA project of the Ministry of Education,
by the National Science Council, Taiwan, R.O.C.
and by the Center for Theoretical Physics
at National Taiwan University.
The work of S.R was supported by DOE grant  
DE-FG02/91ER40688-(Task A).

\end{document}